# Fabrication and Study of Large Area QHE Devices Based on Epitaxial Graphene.

S. Novikov[1], N. Lebedeva[1], K. Pierz[2] and A. Satrapinski[3]

[1]Department of Micro and Nanosciences, Aalto University, Micronova, Tietotie 3, 02150, Espoo, Finland
[2]Physikalisch-Technische Bundesanstalt, Bundesallee 100, 38116 Braunschweig, Germany
[3]MIKES, Tekniikantie 1, FI-02150, Espoo, Finland
Email: novikov@aalto.fi, alexandre.satrapinski@mikes.fi

*ABSTRACT* — **Quantum Hall effect (QHE) devices based on epitaxial graphene films grown on SiC were fabricated and studied for development of the QHE resistance standard. The graphene-metal contacting area in the Hall devices has been improved and fabricated using a double metallization process. The tested devices had an initial carrier concentration of $(0.6 - 10) \cdot 10^{11}$ cm$^{-2}$ and showed half-integer quantum Hall effect at a relatively low (3 T) magnetic field. Application of the photochemical gating method and annealing of the sample provides a convenient way for tuning the carrier density to the optimum value. Precision measurements of the quantum Hall resistance (QHR) in graphene and GaAs devices at moderate magnetic field strengths ($\leq 7$ T) showed a relative agreement within $6 \cdot 10^{-9}$.**

## I. INTRODUCTION

In recent years, significant progress has been achieved in the precision measurement of the quantum Hall effect (QHE) in devices based on epitaxial graphene films grown on SiC [1-4]. Results obtained on large area QHE devices [3] fabricated from epitaxial graphene film showed that the quantization of the Hall resistance in magnetic field as low as 2.5 T is possible due to the application of a photochemical gating [5] which leads to a reduction of the carrier concentration down to $6 \cdot 10^{10}$ cm$^{-2}$. One of the problems related with the fabrication and practical use of such devices for quantum Hall resistance (QHR) measurements is the variation in graphene thickness, leading to inhomogeneity of the film, local variation in carrier density and poor contact resistance to graphene. It is good practice to use high temperatures (1900 ºC - 2000 ºC) [6] for epitaxial graphene growth. In this work we report an improved fabrication technology for epitaxial graphene films grown at lower temperatures (near 1700 ºC), fabrication and modification of the QHE devices with double metallization graphene-metal contacts, and results of experimental studies of the properties of the fabricated graphene QHE devices.

## II. EPITAXIAL GRAPHENE FILM FABRICATION

A set of ten chips with a graphene film was grown on a Si face of 4H-SiC substrates by annealing in Ar ambient at atmospheric pressure and temperatures near 1700 ºC for 5 minutes. AFM measurements show a surface structure with periodical terraces due to a small misorientation of the SiC substrate from the (0001) plane, Fig. 1 a). The height of these terraces was around 0.5 nm. The film thickness was estimated by means of Auger spectroscopy that confirmed the presence of a single layer of graphene before patterning. The number of layers was extracted from the ratio between the Si and C peak using a method described in [7]. Additionally, we have measured Raman spectra, but the interpretation is not straightforward due to stress in the film. It was claimed that "the only unambiguous fingerprint in Raman spectroscopy to identify the number of layers for graphene on SiC(0001) is the line width of the 2D peak" [8]. In our case the FWHM = 40 cm$^{-1}$ of the 2D peak is related to a single graphene layer. According to Auger measurements the total graphene coverage of the SiC surface in different samples varied in the range of 0.7 - 1.05 monolayers. It means that the growth process was completed before the second graphene layer started to grow. However, in a large area epitaxial film the formation of islands with double layer graphene is possible [8]. For details of device fabrication see [3] and [9]. Here we report on the improvements in graphene technology and the fabrication and characterization of the QHE devices.

## III. QHE DEVICE FABRICATION

Patterns for the Hall bars and the contacts were made using laser photolithography with AZ5214 resist. Reactive ion etching in argon-oxygen plasma was applied to remove the graphene layer from uncoated areas.

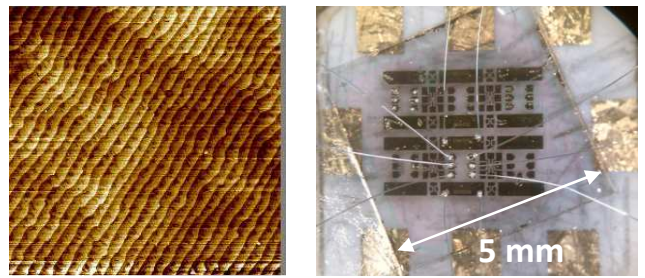

a)   b)

Fig. 1. a) AFM image of the surface of 10 μm x 10 μm epitaxial graphene film and b) photo of $5 \times 5$ mm$^2$ chip with 18 Hall devices.

QHE devices having three Hall contact pairs and different dimensions of the channel (with the largest area 2200 μm × 500 μm) were fabricated on one 5 mm × 5 mm chip. An example of one of the chips with 18 Hall devices is presented in Fig. 1 b). The direction of the current channel was chosen according to an AFM image with the channel directed along the terraces.

The contact resistance at the metal-graphene interface and its stability is a critical property for graphene based devices. The low adhesion of the metallic contacts to the graphene film surface leads to their detachment from the underlying graphene layer upon processing. Formation of carbides on the boundaries of graphene [10] increases the contact resistance. Instability of the contact resistance between metal and graphene layers limits the precision of the measurement.

For fabrication of reliable and low resistance contacts, a two-step metallization process [6] has been used, see Fig. 2. Double metal-graphene-metal contacts were made by e-beam evaporation and lift-off photolithography.

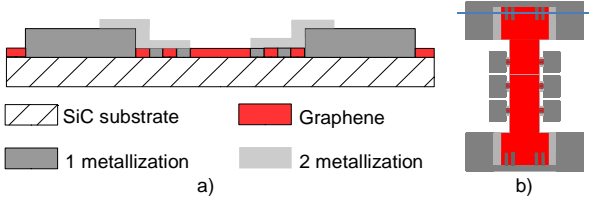

Fig. 2. a) Schematic cross section view of the two-step metallization processes of graphene-metal contact fabrication, b) configuration of one Hall device with strip-like contacts.

In the first steps, Ti/Au (5/50 nm) was used, and in the second, Ti/Au (200/300 nm) metallization was used for reducing the contact resistance. The first metallization was made on the SiC surface areas, where the graphene film was etched. The main advantages of our improved contacts are: a) better adhesion of the contact pads deposited directly to the SiC surface (not to graphene), b) increased fraction of the end-type contacts [11] using stripe shapes, compared to the conventional geometry and c) bonding wires, attached to the contact pads of the first metallization layer, do not damage the graphene film.

## IV. CONTROL OF CARRIER DENSITY

The carrier concentration $n_c$ is determined from the Hall measurement ($R_{xy}$) in low magnetic fields $B$ as $n_c = 1 / [e \cdot (dR_{xy}/dB)]$ and the carrier mobility $\mu = 1/(n_c \cdot R_{sq} \cdot e)$ from the measurement of the square resistance ($R_{sq}$). Two methods for controlling the carrier density were tested in our technological process: exposition in hot air and photochemical gating [5]. Exposition in hot air was performed for one or two hours, respectively, just after fabrication of the Hall bars and the contacts, but before covering with bilayer polymer for the photochemical gaiting method. The values of $n_c$ and $\mu$ were evaluated at 1.5 K and the square resistance was measured at 293 K and at 1.5 K. The parameters of the as-fabricated samples are presented in Table I. Seven chips (#210114, #60214, #210314, #240314, #060314, #190314 and #260514) were kept in air at 120 ºC for different times (0, 1 or 2 hours) in order to introduce additional oxygen related doping centers. During the QHE measurements of these samples it was observed that the reduction of the carrier density was proportional to the time of exposure at 120 ºC before covering with polymer. The samples which were kept in hot air for 2 hours were affected more and the carrier concentration decreased from (0.9 - 1.2) · $10^{12}$ cm$^{-2}$ (for not exposed samples) to (0.6 - 1.3) · $10^{11}$ cm$^{-2}$. This can be explained by hole-doping of the graphene film by adsorbed oxygen molecules [12] and can be used for preliminary tuning of $n_c$, before covering the samples with bilayer polymer.

Table I. Carrier density $n_c$, mobility $\mu$ and sheet resistance $R_{sq}$ of as-fabricated epitaxial QHE devices (before UV illumination)

| N | Chip # | Time of oxidation at 120 ºC | $n_c$ | $\mu$ | $R_{sq}$ | $R_{sq}$ |
|---|--------|---|---|---|---|---|
|   |        |       |           |          | 300 K | 1.5 K |
|   |        | hours | 1/cm$^2$  | cm$^2$/Vs | kΩ | kΩ |
| 1 | 220813 | * | 3.1 10$^{11}$ | 2700 | 10 | 15 |
| 2 | 220813a | ** | - | - | 5.4 | - |
| 3 | 261113-g | * | 9.0 10$^{11}$ | 840 | 4.5-5.6 | 3.7-9.5 |
| 4 | 261113-u | * | 9.0 10$^{10}$ | 1610 | 3.5-5.6 | 4.6 |
| 5 | 210114 | 0 | 8.0 10$^{11}$ | 870 | 9.3-9.8 | 8.9-9.0 |
| 6 | 60214 | 1 | 1.4 10$^{11}$ | 2500 | 10-11 | 15-17 |
| 7 | 210314 | 1 | 5-10 10$^{11}$ | 2950 | 3-6 | 2-4 |
| 8 | 240314 | 0 | 1.5 10$^{12}$ | 1400 | 5-7 | 3-4 |
| 9 | 060314 | 2 | 6.0 10$^{10}$ | 3000 | 14-17 | 20-35 |
| 10 | 190314 | 2 | 6.1 10$^{10}$ | 1280 | 10-13 | 30-60 |
| 11 | 260514 | 2 | 1.3 10$^{11}$ | 3120 | 13.5-14 | 15.5 |

*not controlled, ** was not measured before illumination with UV light; after UV $n_c = 0.64 \cdot 10^{11}$ cm$^{-2}$ and $\mu = 4640$ cm$^2$/Vs

As Table I shows, $n_c$ of the as-fabricated samples varies between 6.0 · $10^{10}$ cm$^{-2}$ – 1.2 · $10^{12}$ cm$^{-2}$ and $\mu$ varies between (970 – 3100) cm$^2$/Vs. The devices from the chips #60214, #210314, #060314, #190314, #260514 (which were exposed in air for 1 – 2 hours before covering with resist) have $n_c$ between (0.6 - 1.3)·$10^{11}$ cm$^{-2}$ and demonstrates a well quantized ν = 2 plateau starting between 3 T and 4 T even without UV illumination. Sheet resistances of these samples are between (5 – 13) kΩ sq at $T$ = 293 K and (4 - 17) kΩ sq at $T$ = 1.5 K. The size of the channel in samples N1 - N6 and in N7 – N9 was 800 μm × 200 μm and 2200 μm × 500 μm accordingly. Samples N5 and N8 (without additional annealing) showed the ν = 2 plateau at $B > 8$ T.

In the photochemical gating method the chips with the Hall devices were covered by two polymers, first by 300 nm of PMMA resist and second by 300 nm of ZEP520A resist. The mechanism of the photochemical gating by UV exposure of the bilayer polymer proceeds via the formation of photo-induced Cl radicals in the top polymer layer. Those radicals act as electron acceptors and control the carrier density of the graphene layer. For tuning $n_c$ to the optimum values between $(0.6 – 3.0)\cdot 10^{11}$ cm$^{-2}$ an UV illumination ($\lambda = 240$ nm, $P = 16$ mW) was used. Our experience shows that the graphene film resistance changes during long time storage, so the illumination of the sample and the adjustment of the carrier density (by measurement of its sheet resistance) were performed just before cooling the sample. The typical cooling rate was about 1 K/min and the sheet resistance was evaluated at $B = 0$ T. An example of the carrier density tuning for one of the samples (#60214) is presented in Fig. 3 and Table II. Fig. 3 shows the changes in position and shape of the plateaus after application of the treatment.

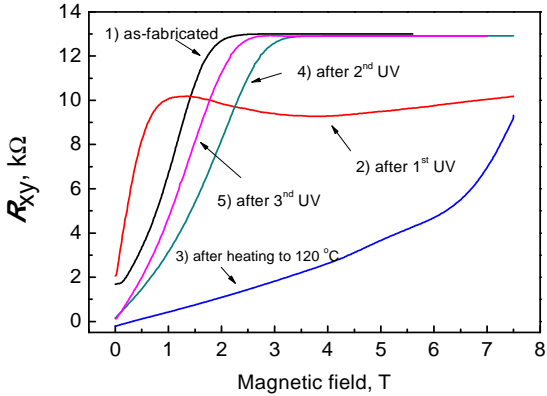

Fig. 3. Tuning of the carrier concentration of the sample (#60214) covered with bilayer polymer by the photochemical method using UV light exposure and annealing at 120 ºC.

Table II. Control of carrier density, mobility and sheet resistance by photochemical gating and heat treatment of epitaxial graphene sample #60214. The sample treatment parameters, date of UV illumination and QHR measurement are presented in the first column.

| Process and Date | $n_c$ (cm$^{-2}$) | $\mu$ (cm$^2$/Vs) | $R_{sq}$ (k$\Omega$) | |
|---|---|---|---|---|
| | $T = 1.5$ K | 1.5 K | 300 K | 1.5 K |
| as fabricated (21.03.2014) | $1.4\cdot 10^{11}$ | 2500 | 10 | 20 |
| 1$^{st}$ UV illumination (31.03.2014) | $7.8\cdot 10^{10}$ | 1830 | 14 | 44 |
| Heating at 120 ºC (1.04.2014) | $1.0\cdot 10^{12}$ | 1470 | 3.3 | 4.3 |
| 2$^{nd}$ UV illumination (2.04.2014) | $3.5\cdot 10^{11}$ | 920 | 9 | 14 |
| 3$^{rd}$ UV illumination (11.07.2014) | $1.3\cdot 10^{11}$ | 2700 | 11 | 17 |

After the first illumination (curve 2) the carrier density decreased from $1.4\cdot 10^{11}$ cm$^{-2}$ to $7.8\cdot 10^{10}$ cm$^{-2}$ but the $\nu = 2$ plateau became asymmetric and tilted and strongly deviated from the expected value of $R_K/2 \approx 12.906$ k$\Omega$, where $R_K$ is the von Klitzing constant. The square resistance increased to 40 k$\Omega$ at 1.5 K. In the next step, heating at 120 ºC increased $n_c$ to $1.0\cdot 10^{12}$ cm$^{-2}$. Subsequent UV illumination steps returned $n_c$ first to $3.5\cdot 10^{11}$ cm$^{-2}$ and then to $1.3\cdot 10^{11}$ cm$^{-2}$. Deviation from the nominal value of $R_K/2$ can be observed due to incomplete quantization of the 2DEG in QHE devices caused by local dissipation. Measurement of the longitudinal resistance $R_{xx}$ is used to define the conditions (biasing current and temperature) at which $R_{xx}$ is sufficiently small for precision measurements [13]. This confirms the quantization of the 2DEG. In the tested sample, low (at the level of $10^{-4}$ $\Omega$) but non equal values of $R_{xx}$ were observed in different areas of the channel, which indicates that in large area epitaxial films defects and domains with various thicknesses are present [14,15,16]. The existence of areas with bilayer inclusions can act as equipotential shorts for edge currents [16] and can be the reason for inhomogeneous carrier density distribution and deviations of $R_H(2)$ from the expected value of $R_K/2$.

## V. LONGITUDINAL RESISTANCE MEASUREMENT

The longitudinal resistance, $R_{xx}$, in a magnetic field corresponding to the Hall plateau has been measured before the precision QHE measurements. Current dependence measurements of $R_{xx}$ have been performed on different Hall contacts along the channel.

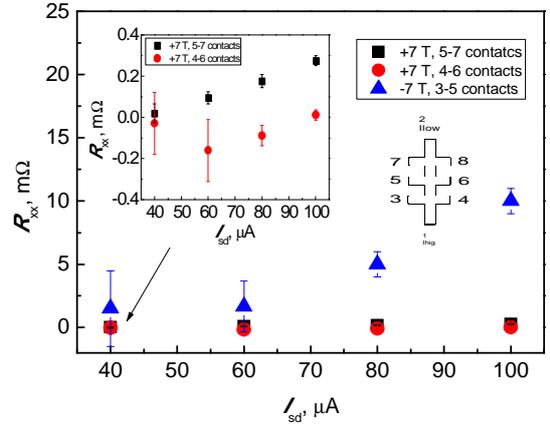

Fig. 4. Current dependence of longitudinal resistance, $R_{xx}$, in sample #220813, measured using different contact pairs and in positive (black, red) and negative (blue) magnetic field direction. Inset shows $R_{xx}$ for two contacts pairs on a scale with higher resolution. For currents less than 80 µA, $R_{xx}$ is less than 0.2 m$\Omega$ for those contact pairs in positive magnetic field.

These measurements showed a relatively large difference in $R_{xx}$ in different areas of the channel. An example of the current dependence of $R_{xx}$ in the measurement using sample #220813, on the contacts 4 - 6 and 5 – 7 is presented in Fig. 4. Longitudinal resistance was measured using a nanovoltmeter (Keithley 2182) and applying dc currents with reversing polarity and measuring the voltage on the corresponding contacts. As can be seen in the inset of Fig 4, $R_{xx}$ measured at

$B = +7$ T on different contacts with applied currents up to 100 µA increased from 30 µΩ to about 0.3 mΩ with increasing current. Variations of $R_{xx}$ on the contacts 4 - 6 and 5 - 7 with $I_{sd}$ up to 100 µA indicate possible inhomogeneity of the carrier density along the channel. Longitudinal resistance $R_{xx}$ increased up to 10 mΩ for contacts 3 – 5. The current dependence of $R_{xx}$ of another graphene sample, #220813a, is presented in Fig. 5, for $B = 6$ T (black line) and for $B = 7$ T (red line). The sample was illuminated with UV light and the carrier density was tuned to $0.6 \cdot 10^{11}$ cm$^{-2}$ and µ = 4640 cm$^2$/Vs. $R_{xx}$ was below 2 mΩ at currents lower than 80 µA, when the applied magnetic field was 6 T or 7 T. With increasing $I_{sd}$, $R_{xx}$ increased and became larger (up to 10 mΩ) for a lower magnetic field of 6 T.

VI. QHE MEASUREMENT AT DIFFERENT HALL BARS

An example of the general dependence of the Hall resistance versus magnetic field measured on an epitaxial graphene QHE device (#220813) using three Hall contact pairs is shown in Fig. 6. It is seen that for all of these Hall contact pairs, the ν = 2 plateau begins at a magnetic field of about 4 T. Although the sample was not illuminated with UV light, its initial carrier density varied between $(1.6 - 3.1) \cdot 10^{11}$ cm$^{-2}$ and the mobility was between 1400 and 2700 cm$^2$/Vs.

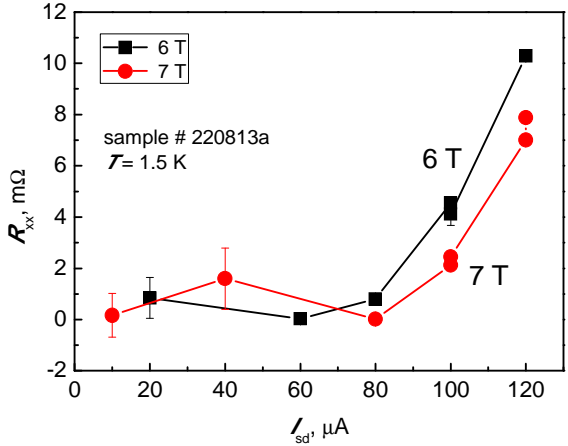

Fig. 5. The current dependence of $R_{xx}$ (at $T = 1.5$ K) for $B = 6$ T and $B = 7$ T in graphene sample #220813a. The longitudinal resistance is less than 2 mΩ for $I_{sd}$ below 80 µA.

Precision QHE measurements using a low frequency Cryogenic Current Comparator (CCC) Resistance Bridge [17] have been performed with two graphene samples; #220813 and #260514 (both with channel size 800 µm × 200 µm) and with a custom made low density GaAs device [18]. In Fig. 7 the results of a comparative measurement of a 100 Ω resistance standard using graphene (#220813) and GaAs QHE devices are presented.

Figure 8 shows the calibration history of the 100 Ω standard resistor. The low carrier density GaAs device fabricated at PTB (G-1) has been used as a reference in QHR measurements with the stable 100 Ω standard.

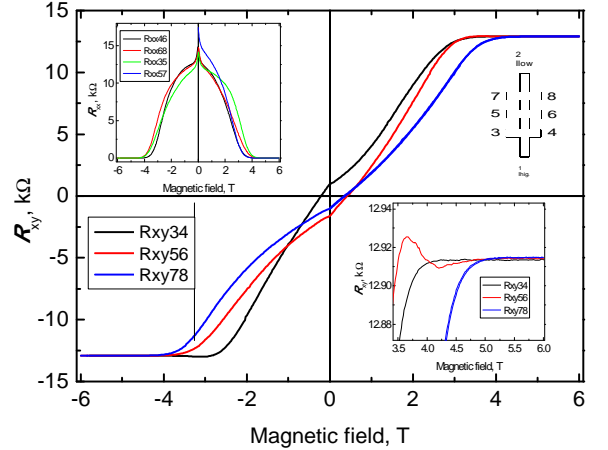

Fig. 6. General dependence of $R_{xy}$ versus magnetic field at three Hall contact pairs, measured at $T = 1.5$ K with 10 µA current. The inset in the lower right corner shows the Hall resistance between 4.5 T and 6 T. The inset in the upper left part shows longitudinal resistance $R_{xx}$ versus magnetic field on different contact pairs.

Measurements with the GaAs sample showed that it has a well quantized ν = 2 plateau between 5.5 T – 5.8 T, at $T = 1.5$ K, and $I_{sd} = 10$ µA.

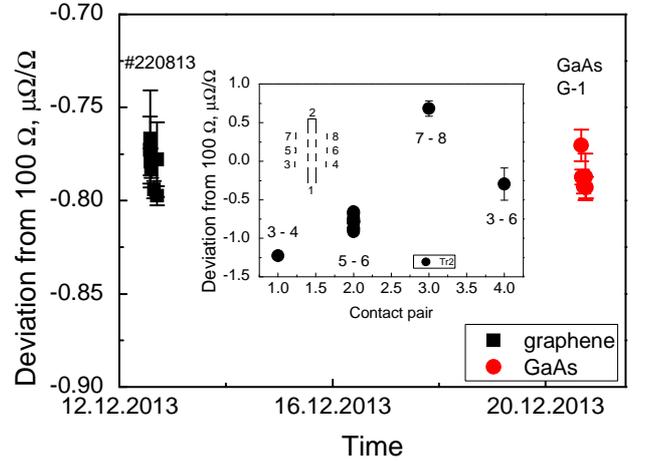

Fig. 7. Comparative QHE measurements of a 100 Ω resistance standard performed using middle Hall contact pairs (5-6) of epitaxial graphene and GaAs samples. The inset shows the results of the QHE measurements performed using the graphene device at different Hall contact pairs. Error bars are 1 σ statistical uncertainty.

The QHR measurements with the GaAs device have been performed against $R_H(2)$ at filling factor ν = 2, B = 5.70 T, T = 1.5 K, and with $I_{sd}$ = 26 µA (rms).

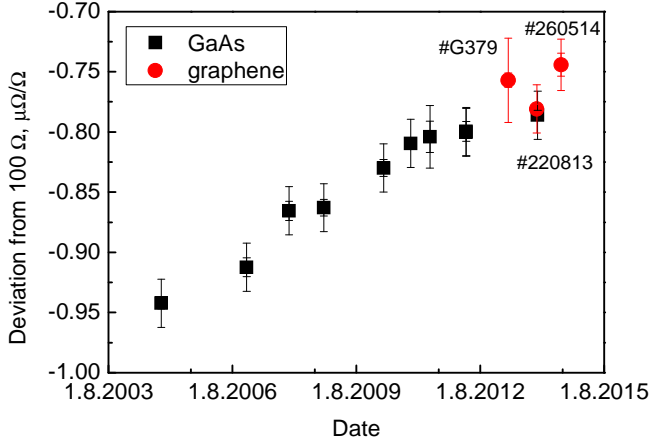

Fig. 8. Results of the measurements of the 100 Ω standard against the GaAs samples and against the graphene samples, (#G379, #220813 and #260514). Error bars are combined expanded (k=2) uncertainty.

QHE measurements of the graphene samples have been performed on the ν = 2 plateau, at B = +7.0 T, T = 1.5 K, with $I_{sd}$ = 33 µA (rms) and on three different Hall contact pairs. A two terminal measurement made on graphene, at Hall contact pairs (1 - 2, 5 - 6 and 3 - 4), showed the expected resistance of 12.91 kΩ, but an increased resistance at contact pair (7 – 8). The value of $R_H(2)$ at contacts 7 - 8 deviates from the expected $R_K/2$ value by about $0.6 \cdot 10^{-6}$, see inset in Fig 7. The relative difference in the measurement of the 100 Ω standard by a CCC Bridge between the values obtained using GaAs and graphene samples at the middle Hall contact pair is $0.005 \cdot 10^{-6}$ with combined statistical uncertainty of $0.006 \cdot 10^{-6}$ (k = 1). The inset in Fig. 7 shows the variations of the results obtained at different Hall contact pairs using this graphene device. These results also indicate that there is inhomogeneity and non-uniformity of the graphene films along the channel leading to a contact pair dependence of the measured $R_{xy}$.

One of the criteria to demonstrate the quality and applicability of the QHE samples is a low contact resistance in current and Hall potential terminals. An example of the influence of an increased contact resistance in Hall contact pairs can be seen in the results of the QHE measurement performed with graphene sample #260514. This sample is on a 2.5 mm × 2.5 mm chip and has a channel size of 800 µm × 200 µm. Results of the QHR measurements are presented in Fig. 9 and in Table III. Before the precision measurement, two terminal measurements of the Hall resistance (made with a hand-held DVM) at the ν = 2 plateau and B = +7.0 T were performed at three contact pairs, see Table III. For this sample we measured $R_{xx} \leq 15$ mΩ using contacts 14h -15h and 10h -

13h at B = 7.0 T, T = 1.5 K and for $I_{sd}$ < 80 µA. Precision QHR measurements have been performed via a CCC Bridge with the same stable 100 Ω standard against $R_H(2)$ at B = 7.0 T, T = 1.5 K with $I_{sd}$ = 26 µA (rms) at three Hall contact pairs.

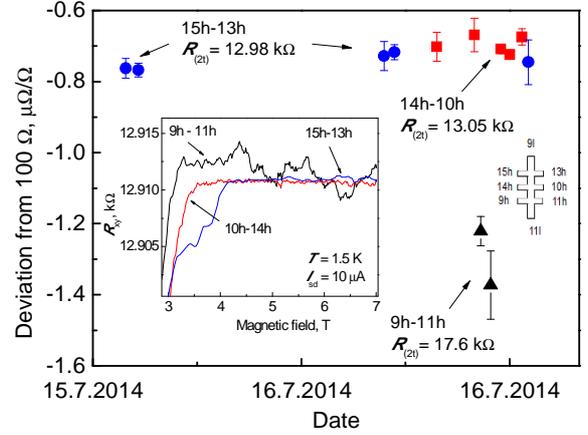

Fig. 9. Deviation from nominal value of the 100 Ω standard resistor, measured using the ν = 2 plateau of the QHR in the graphene sample #260514, at three Hall contact pairs. Inset shows plateau flatness measured at the same Hall contact pairs. Increased two terminal resistance and non-flatness of the plateau measured at 9h - 11h contact pair correlates with the deviation of $R_{xy}(2)$ measured at this contact pair.

Table III. The results of two-terminal resistance measurements at three different Hall contact pairs and the deviation from the expected value of the 100 Ω standard resistor, measured at the same contact pair.

| Contact pair | Two terminal Hall resistance | Deviation from expected value of 100 Ω | $u_c$ (k = 1) |
|---|---|---|---|
| | kΩ | µΩ/Ω | µΩ/Ω |
| 15h – 13h | 12.98 | 0.014 | 0.015 |
| 14h – 10h | 13.05 | 0.063 | 0.015 |
| 9h – 11h | 17.60 | -0.538 | 0.08 |

There is a large relative deviation ($-0.54 \cdot 10^{-6}$) from the expected value of the 100 Ω resistor, measured on a "high contact resistance" (9h-11h) Hall pair (see Fig 9, black triangles), and Table III. The inset in Fig. 9 shows the shapes of ν = 2 plateau measured at these Hall contact pairs, at T = 1.5 K and with $I_{sd}$ = 10 µA. It is seen that at a "high contact resistance" Hall pair the ν = 2 plateau starts from 3 T, however, it is not flat and has quite large fluctuations (several ohms). The observed correlation between increased two-terminal resistance and fluctuations on the Hall plateau suggests that the increased contact resistance can be one of the reasons of the observed deviations in some of the QHR measurements using epitaxial graphene [3-4].

Figure 8 shows a 10 year calibration history of the 100 Ω standard resistor that was used in the experiments presented in Figures 7 and 9. Measurement results against $R_H(2)$ of GaAs and three graphene devices (#G379, #220813 and #260514, red squares) are presented, too.

VIII. CONCLUSION

A set of QHE devices based on epitaxial graphene on 4H-SiC substrate were fabricated. The Magneto-transport measurements of the devices having carrier density within (1 - 3) · $10^{-11}$ cm$^{-2}$ showed half-integer quantum Hall effect with a ν = 2 plateau starting at a relatively low magnetic field (3 T). Inhomogeneity of the carrier density distribution and variation of the carrier mobility in different areas of the device were the main factors limiting the precision measurements at the Hall contact pairs along the channel. In spite of partial variations of the carrier density, the comparative measurements performed with the CCC Bridge of a 100 Ω standard resistor against $R_H(2)$ on two tested graphene samples (at the middle Hall contact pairs, B = 7 T, T = 1.5 K and $I_{sd}$ = 33 µA) and on a GaAs sample, showed an agreement within a relative uncertainty of (6 - 10) · $10^{-9}$.

ACKNOWLEDGEMENT

This study was supported in part by EMRP project SIB51, "GraphOhm". The EMRP is jointly funded by the EMRP participating countries within EURAMET and the European Union. We thank A. Manninen and F. Ahlers for useful discussions and C. Richmond for assistance in preparation of the article.